\documentclass[aps,prb,twocolumn,superscriptaddress,amsmath,
               amssymb,footinbib]{revtex4-1}
\usepackage{newtxtext,newtxmath}
\usepackage{graphicx}
\usepackage[dvipsnames]{xcolor}
\usepackage{appendix}
\usepackage[colorlinks=true,citecolor=blue,breaklinks=true,
            linkcolor=red,linktocpage=true,urlcolor=blue,
            pagebackref=false]{hyperref}

\newcommand{\ev}[1]{\ensuremath{\langle #1 \rangle}}
\newcommand{\abs}[1]{\ensuremath{|#1|}}

\newcommand{\braketop}[3]{\ensuremath{\langle #1|#2|#3\rangle}}

\newcommand{\ket}[1]{\ensuremath{|#1\rangle}}
\newcommand{\up}{{\uparrow}}
\newcommand{\down}{{\downarrow}}

\renewcommand{\paragraph}[1]{{\par\it #1.---}\ignorespaces}

\begin{document}
\title{Single-photon pump by Cooper-pair splitting}

\author{Mattia Mantovani}
\affiliation{Fachbereich Physik, Universit\"at Konstanz, D-78457 Konstanz,
             Germany}
\author{Wolfgang Belzig}
\affiliation{Fachbereich Physik, Universit\"at Konstanz, D-78457 Konstanz,
	Germany}
\author{Gianluca Rastelli}
\affiliation{Fachbereich Physik, Universit\"at Konstanz, D-78457 Konstanz,
             Germany} \affiliation{Zukunftskolleg, Universit\"at Konstanz,
             D-78457 Konstanz, Germany}
\author{Robert Hussein}
\affiliation{Fachbereich Physik, Universit\"at Konstanz, D-78457 Konstanz,
             Germany}
\date{\today}

\begin{abstract}
Hybrid quantum dot-oscillator systems have become attractive platforms to
inspect quantum coherence effects at the nanoscale. Here, we investigate a
Cooper-pair splitter setup consisting of two quantum dots, each linearly coupled
to a local resonator. The latter can be realized either by a microwave cavity or
a nanomechanical resonator. Focusing on the subgap regime, we demonstrate that
cross-Andreev reflection, through which Cooper pairs are split into both dots,
can induce nonlocal correlations between the two resonators. Harnessing these 
correlations allows to establish and control a nonlocal photon transfer between them. 
The proposed scheme can act as a photonic valve with single-photon accuracy, 
with potential applications for quantum heat engines and refrigerators 
involving mesoscopic resonators.
\end{abstract}

\maketitle

\section{Introduction}
Nonlocality~\cite{Buhrman2010,Brunner2014} and quantum correlations~\cite{Eckert2002} are at the heart
of many quantum technologies.~\cite{Monroe2002,Ladd2010,Degen2017} In hybrid quantum-dot devices, Cooper pairs
are a source of correlated electrons and their nonlocal splitting has experimentally~\cite{Hofstetter2009, Herrmann2010, Hofstetter2011, Das2012, Schindele2012,
Schindele2014, Fulop2014, Fulop2015, Tan2015, Borzenets2016, Baba2018} and theoretically~\cite{Recher2001, Chevallier2011,
Rech2012, Scherubl2014, Trocha2015, Nigg2015, Schroer2015, Probst2016, Dominguez2016, Amitai2016, Hussein2016,
Wrzesniewski2017a, Hussein2017, Wrzesniewski2017,Walldorf2018,Bocian2018} drawn much attention
over the last few years. In particular, the nonlocal breaking of the particle-hole symmetry in such Cooper-pair splitters (CPSs)
gives rise to peculiar thermoelectric effects.~\cite{Cao2015,Sanchez2018,Hussein2019,Kirsanov2019}
On the other hand, mesoscopic  cavity quantum electrodynamics (cQED) devices~\cite{Childress2004,Cottet2017} are excellent tools for correlating few-level systems over a
distance.~\cite{Delbecq2013,Lambert2013,Deng2015,Deng2016,Ramirez-Munoz2018,Trif2019} Such cQED devices
have applications in the readout of charge,~\cite{Frey2012,Cottet2012,Viennot2014,Bruhat2016,Viennot2016,Stockklauser2017,Burkard2019}
spin,~\cite{Petersson2012,Viennot2015,Benito2017,Zhu2018,Mi2018} and valley-orbit
states,~\cite{Kohler2018,Mi2018a} as well as few-phonon manipulation when mechanical modes can be cooled
close to the ground state.~\cite{OConnell2010,Piovano2011,Teufel2011,Stadler2016,Stadler2017,Clark2017}
A mechanism which induces nonlocal photon or phonon correlations through Cooper pair transport, implemented in a hybrid setup, 
bridges the gap between the study of heat flows in quantum-dot-based \cite{Cao2015,Erdman2017,Erdman2018,Sanchez2018,Hussein2019} 
and circuit-QED devices.~\cite{Hofer2016,Ronzani2018,Dambach2019}

\begin{figure}[t]
    \includegraphics[]{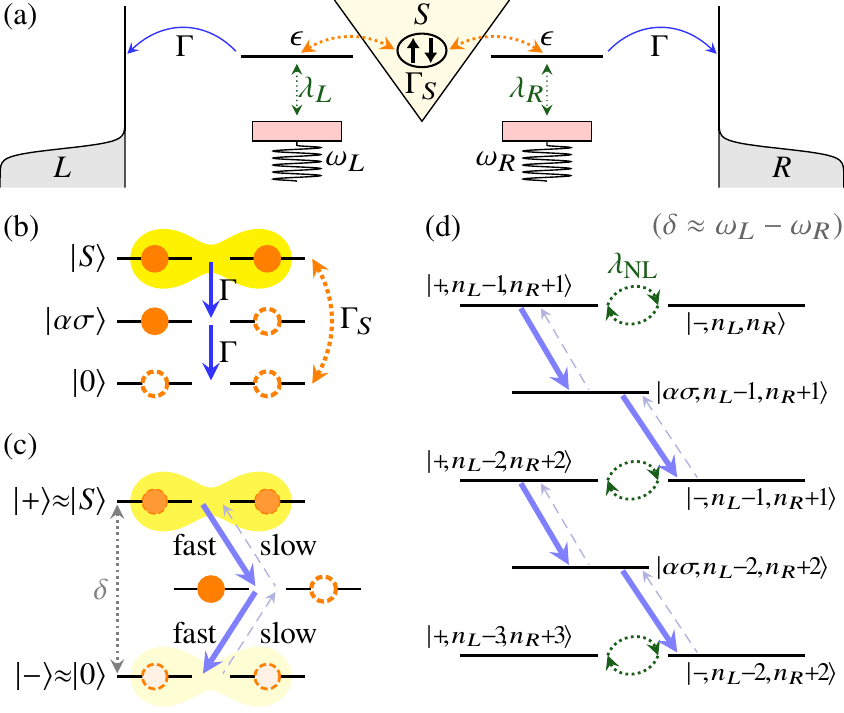}
    \caption{\label{fig:1} (a) Cooper-pair splitter consisting of two quantum
             dots coupled to a common superconductor ($S$) and two normal-metal
             contacts ($\alpha=L,R$). Each dot is capacitively coupled to a local resonator with frequency $\omega_\alpha$. 
             (b) At large bias voltage, incoherent
             tunneling events at rate $\Gamma$ lead to a decay of the singlet state, $|S\rangle$, via a singly-occupied one, $|\alpha\sigma \rangle$ ($\sigma = \uparrow, \downarrow$), to the empty state, $|0\rangle$, whereby $|0\rangle$ and $|S\rangle$ are coherently coupled with amplitude $\Gamma_S$.  
             (c) The latter coupling leads to the formation of hybridized $|\pm\rangle$ states of energy splitting $\delta$. For weakly hybridized states $|0\rangle$ and $|S\rangle$, the transitions $|\pm\rangle \leftrightarrow
             |\alpha\sigma\rangle$ are strongly asymmetric.
             (d) Photon transfer cycle occurring around the resonance, $\delta \approx \omega_L - \omega_R$, with the effective 
             coupling strength $\lambda_{\mathrm{NL}}$.}
\end{figure}
In this work, we consider a CPS in a double-quantum-dot setup with each dot
linearly coupled to a local resonator, constituted by either a microwave cavity
\cite{Viennot2015, Stockklauser2015, Bruhat2016, Mi2017, Stockklauser2017,
Li2018,Cubaynes2019} or a mechanical oscillator,
\cite{Benyamini2014,Okazaki2016,Deng2016,Wen2019,Urgell2019} see 
Fig.~\ref{fig:1}(a). 
We demonstrate that this system is a platform to obtain full control on the
heat and photon exchange of two originally uncoupled cavities.
This induced coupling arises from the proximity between the dots and the superconducting 
lead, and has a purely nonlocal origin due to cross-Andreev reflection.
Subsequent, we discuss the underlying physical mechanism following the lines of Ref.~\onlinecite{Rastelli2019}, where a single-quantum-dot system in the single-atom lasing regime has been investigated.

For large intradot Coulomb interactions, $U$, and superconducting gap, $|\Delta|\to\infty$, the proximity of the superconductor causes a nonlocal splitting (and recombination) of Cooper pairs into both dots with the pairing amplitude $\Gamma_S>0$. The corresponding Andreev bound states $|\pm\rangle$
are a coherent superposition of the dots' singlet, $|S\rangle$, and empty state, |$0\rangle$. The dots are further tunnel-coupled to normal contacts, which are largely negative-voltage-biased with respect to the chemical potential $\mu_S = 0$ of the superconductor. In this configuration, due to single-electron
tunneling, the singlet state decays at
rate $\Gamma$ into a singly-occupied state, $|\alpha \sigma \rangle$ ($\alpha = L,R$ and $\sigma = \uparrow, \downarrow$) and further
into the empty state, see Fig.~\ref{fig:1}(b). For large dot onsite energies $\epsilon\gtrsim\Gamma_S$, the charge hybridization is
weak ($|+\rangle\approx|S\rangle$, $|-\rangle\approx|0\rangle$), and the transitions
$|+\rangle \rightarrow |\alpha\sigma \rangle$ and $|\alpha \sigma \rangle
\rightarrow |-\rangle$ are faster than the opposite processes,~\cite{Rastelli2019} see Fig.~\ref{fig:1}(c).
This asymmetry in the relaxation ultimately explains how to pump or absorb energy within a single mode, and how to transfer photons between the cavities. In the latter case, when the energy
splitting $\delta$ between the Andreev bound states is close to the
difference of the cavity frequencies, the relevant level structure of the uncoupled system
is summarized in Fig.~\ref{fig:1}(d). We show below that the
effective interaction couples the states $|+,
n_L\!-\! 1, n_R \!+\!1\rangle$ and $|-,
n_L, n_R\rangle$, where $n_\alpha$ indicates the Fock number in the
resonator $\alpha$. An electron tunneling event favours 
transitions
\mbox{$|+\rangle \rightarrow |\alpha\sigma\rangle \rightarrow |-\rangle$}
conserving the photon number. When the system reaches the state $|-\rangle\approx|0\rangle$, this coherent cycle restarts. When the system is in $|+\rangle$, it can again decay. During each cycle, a boson is effectively transferred from the left to the right cavity. Since the two cavities are not isolated, but naturally coupled to external baths, a steady heat flow is eventually established between the cavities.

The effect discussed above refers to a single operation point of the system. More generally, using a master equation approach, we show that the interaction between the CPS and the two resonators opens a rich set of inelastic resonant channels for the electron current through the dots, involving either absorption/emission of photons from a local cavity or nonlocal transfer processes. By tuning $\epsilon$ to match these resonances, the CPS acts as a switch allowing the  manipulation of heat between the resonators.
Each resonant process can be captured with good approximation by an effective Hamiltonian which is valid close to the resonance and generalizes the mechanism described above.

This work is structured as follows. After introducing our model and the employed master equation in Sec.~\ref{sec:model}, we provide therein an effective Hamiltonian describing local and nonlocal transport processes. 
In Section~\ref{sec:local}, we discuss the possibility of simultaneous cooling (and heating) of the resonators. Section~\ref{sec:nonlocal} is dedicated to the nonlocal photon transfer between them, and in Sec.~\ref{sec:efficiency} we analyze the efficiency of this transfer. Finally, we draw our conclusions in Sec.~\ref{sec:conclusions}.

\section{Cooper-pair splitter coupled to resonators}
\label{sec:model}
We consider the effective model for two single-level quantum dots proximized by
a $s$-wave superconductor, and each linearly coupled to a local harmonic
oscillator. For large intradot Coulomb interaction, $U\gg|\epsilon|$, the subgap physics of the system is described by the effective Hamiltonian
\cite{Hussein2016,Rozhkov2000,Meng2009,Eldridge2010,Braggio2011,Droste2012,
Sothmann2014,Weiss2017,Walldorf2018}
\begin{equation}
	\label{eq:hamiltonian}
	\begin{split}
	H ={ }&
		\sum_{\alpha\sigma}\epsilon N_{\alpha\sigma} -
		\frac{\Gamma_S}{2}(
			d^\dag_{R\up} d^\dag_{L\down} -d^\dag_{R\down} d^\dag_{L\up} +
            \mathrm{H.c.}
		) \\
	+&\sum_\alpha \omega_\alpha b^\dag_\alpha b_\alpha
	+\sum_{\alpha,\sigma}\lambda_\alpha (b_\alpha + b^\dag_\alpha)
    N_{\alpha\sigma},
	\end{split}
\end{equation}
where $\hbar=1$.
Here, $d_{\alpha\sigma}$ is the fermionic annihilation operator for a
spin-$\sigma$ electron in dot $\alpha$, with the corresponding number operator $N_{\alpha\sigma}$ and
onsite energy $\epsilon$. The interaction of the dot with the
$\alpha$-oscillator of frequency $\omega_\alpha$ and corresponding bosonic field
$b_\alpha$ is realized through the charge term, with coupling constant
$\lambda_\alpha$.
The relevant subspace of the electronic subsystem is spanned by six states: The
empty state $\ket{0}$, the four singly-occupied states $\ket{\alpha \sigma}=
d^\dag_{\alpha \sigma}\ket{0}$ and the singlet state 
$\ket{S}=\frac{1}{\sqrt{2}}
(d^\dag_{R\up} d^\dag_{L\down} -d^\dag_{R\down} d^\dag_{L\up})\ket{0}$. 
Triplet states 
and doubly-occupied states are inaccessible due to 
large negative voltages, see Fig.~\ref{fig:1}(a), and large intradot Coulomb repulsion.  
Finally, in the subgap regime, the superconductor can only pump 
Cooper pairs, which are in the singlet state. 
The states $|0\rangle$ and $|S\rangle$ are hybridized due 
to
the $\Gamma_S$-term, yielding the Andreev states
$|+\rangle = \cos (\theta/2) |0 \rangle + \sin (\theta/2) |S \rangle$ and $|-
\rangle = - \sin (\theta/2) |0 \rangle + \cos(\theta/2) | S \rangle$, with the
mixing angle
$\theta = \arctan [\Gamma_S/(\sqrt{2} \epsilon)]$. We 
denote
their energy splitting by $\delta = \sqrt{\smash[b]{4 \epsilon^2 + 2
		\Gamma_S^2}}$.

Electron tunneling into the normal leads and dissipation for the
resonators can be treated in the sequential-tunneling regime to lowest order in
perturbation theory, assuming  small dot-lead tunneling rates, $\Gamma \ll
\Gamma_S, k_B T$, and large quality factors $Q_\alpha =\omega_\alpha/\kappa_\alpha$
for the resonators, i.e. $\kappa_\alpha \ll\omega_\alpha, k_B T$. Here,
$\kappa_\alpha$ is the decay rate for the $\alpha$-resonator and $T$ is the temperature of the fermionic and bosonic reservoirs.
The fermionic and bosonic transition rates
between two eigenstates $|i\rangle$ and $|j\rangle$ of
Hamiltonian~\eqref{eq:hamiltonian} are given by Fermi's golden
rule,~\cite{Benenti2017}
\begin{align}
	w^{\alpha, s}_{\mathrm{el}, j \leftarrow i} &=
		\Gamma f^{(s)}_\alpha (sE_{ji}) \sum_{\sigma}\abs{\braketop{j}{ 
		d^{(s)}_{\alpha \sigma} }{i}}^2 	\label{eq:fermionic_rates}, \\
	w^{\alpha, s}_{\mathrm{ph}, j \leftarrow i} &=
		s \kappa_\alpha n_{B}(E_{ji}) \abs{\braketop{j }{ b^{(s)}_{\alpha} 
		}{i}}^2 \label{eq:bosonic_rates},
\end{align}
with \mbox{$f^{(s)}_\alpha(x) = \{\exp[s(x-\mu_\alpha)/k_BT] + 1\}^{-1}$}
the generalized Fermi function ($s=\pm$) at chemical potential $\mu_\alpha$, and
\mbox{$n_{B}(x) = [\exp(x/k_BT) - 1]^{-1}$} the Bose function. \mbox{$E_{ji} \equiv E_j -
E_i$} denotes the energy difference between two eigenstates. We use the notation
$d^{(-)}_{\alpha\sigma}$ ($d^{(+)} _{\alpha\sigma}$) for fermionic
annihilation (creation) operators, and correspondingly $b^{(\pm)}_\alpha$ for the bosonic ones. The populations $P_i$ of the system eigenstates 
obey a Pauli-type master equation of
the form~\cite{Sauret2004,Governale2008,Hussein2016}
\begin{equation} \dot{P}_i = \sum_j w_{i \leftarrow j} P_j - \sum_j w_{j
  \leftarrow i} P_i    \label{eq:master_equation},
\end{equation}
which admits a stationary solution given by $P^\mathrm{st}_i$. The total rates
entering Eq.~\eqref{eq:master_equation} are given by $w_{j \leftarrow i} =
\sum_{\alpha, s}(w^{\alpha, s}_{\mathrm{el}, j \leftarrow i} +
w^{\alpha,s}_{\mathrm{ph}, j \leftarrow i}).$
As mentioned before, we assume the chemical potentials of the normal leads 
$\mu_\alpha = -eV$ to be largely negative-biased, i.e., $U,\abs{\Delta}\gg eV 
\gg k_B T, \epsilon, \Gamma_S$, with $V>0$ and $e>0$ denoting the applied 
voltage and the electron charge, respectively. In 
this regime, the electrons flow unidirectionally from the superconductor via 
the quantum dots into the leads; the temperature of the normal leads becomes 
irrelevant, and the rates $w_{\mathrm{el}, j \leftarrow i}^{\alpha, +}$ 
vanish. Under these assumptions, the stationary electron current through lead 
$\alpha$ is simply given by $I_\alpha= e \Gamma \sum_{\sigma} \langle 
N_{\alpha\sigma} \rangle$.
For a symmetric configuration, as assumed here, both stationary currents 
coincide, $I_L=I_R$.
To evaluate the stationary current and the other relevant quantities, we 
diagonalize 
numerically Hamiltonian~\eqref{eq:hamiltonian}, and build the transition-rate 
matrices appearing in Eq.~\eqref{eq:master_equation}. The stationary populations, $P_i^{\mathrm{st}}$, are then found by solving
the system of Eqs.~\eqref{eq:master_equation} for $\dot{P}_i = 0$.

%
In order to explain our numerical results, we 
perform the Lang-Firsov polaron 
transformation to 
Hamiltonian~\eqref{eq:hamiltonian}.~\cite{Lang1962,Brandes1999,Brandes2005} 
For an operator $O$, we define the unitary
transformation $\bar{O} = e^{\xi} O e^{- \xi}$, with $\xi =
\sum_{\alpha\sigma} \Pi_\alpha N_{\alpha\sigma}$ and $\Pi_\alpha =  
(\lambda_\alpha
/\omega_\alpha) (b^\dagger_\alpha - b_\alpha)$. The 
polaron-transformed Hamiltonian reads then
\begin{equation}
\label{eq:polaron_transformed_hamiltonian}
\bar{H} = \sum_{\alpha\sigma}\bar{\epsilon}_{\alpha} N_{\alpha\sigma} - 
\frac{\Gamma_S}{\sqrt{2}} (|S \rangle \langle 0 | X + |0 \rangle \langle S | 
X^\dagger) + \sum_\alpha \omega_\alpha b_\alpha^\dagger b_\alpha,
\end{equation}
with \mbox{$\bar{\epsilon}_{\alpha}
= \epsilon - \lambda_\alpha^2/\omega_\alpha$} and \mbox{$X = 
\exp(\sum_{\alpha} \Pi_{\alpha})$}.\footnote{Notice that the polaron 
transformation also gives rise to a modification of the tunneling Hamiltonian, 
which describes inelastic dot-lead tunneling events scaling with $\Gamma 
(\lambda_\alpha/\omega_\alpha)^2$. However, in our operative regime, such 
processes can be neglected since $\Gamma \ll \Gamma_S$ and $\lambda_\alpha \ll 
\omega_\alpha$ ensure $\Gamma 
(\lambda_\alpha/\omega_\alpha)^2 \ll \Gamma  \ll \Gamma_S$. }
Equation~\eqref{eq:polaron_transformed_hamiltonian} contains a
transverse charge-resonator interaction term to all orders in the couplings
$\lambda_\alpha$. Intriguingly, this coupling has a purely nonlocal origin
stemming from the cross-Andreev reflection. By expanding $X$ in powers of 
$\Pi\equiv\sum_\alpha\Pi_\alpha$ 
assuming small couplings $\lambda_\alpha \ll \omega_\alpha$, and moving to the 
interaction 
picture with respect to the noninteracting Hamiltonian, we can identify 
a family of resonant conditions given by
\begin{equation}
\label{eq:resonances}
\bar{\delta} \approx |p \omega_L \pm q \omega_R|,
\end{equation}
with $p, q$ nonnegative integers, as discussed in Appendix~\ref{app:app}.
Here, $\bar{\delta} =\sqrt{\smash[b]{4\bar{\epsilon}^2+ 2 
\Gamma_S^2}} $ is the renormalized energy splitting of the Andreev states due to the polaron shift, 
with $\bar{\epsilon} = \epsilon -  \sum_\alpha 
\frac{\lambda_\alpha^2 }{2\omega_\alpha}$.  The renormalized mixing angle reads $\bar{\theta} = 
\arctan[\Gamma_S/(\sqrt{2}\bar{\epsilon}) ]$.
Around the conditions stated in Eq.~\eqref{eq:resonances}, a rotating-wave approximation yields an
effective interaction of order $p+q$ in the couplings $\lambda_\alpha$. 
Hereafter, we discuss in detail the resonances at $\bar\delta = \omega_L = 
\omega_R$ and $\bar\delta = \omega_L - \omega_R$ corresponding to one- and
two-photon processes, respectively. They can be fully addressed by expanding $X$ up to second order in 
$\lambda_\alpha/\omega_\alpha$ and subsequently performing a rotating-wave approximation,
see Appendix~\ref{app:app}.

\section{Simultaneous cooling and heating}
\label{sec:local}
\begin{figure}[t]
	\includegraphics[]{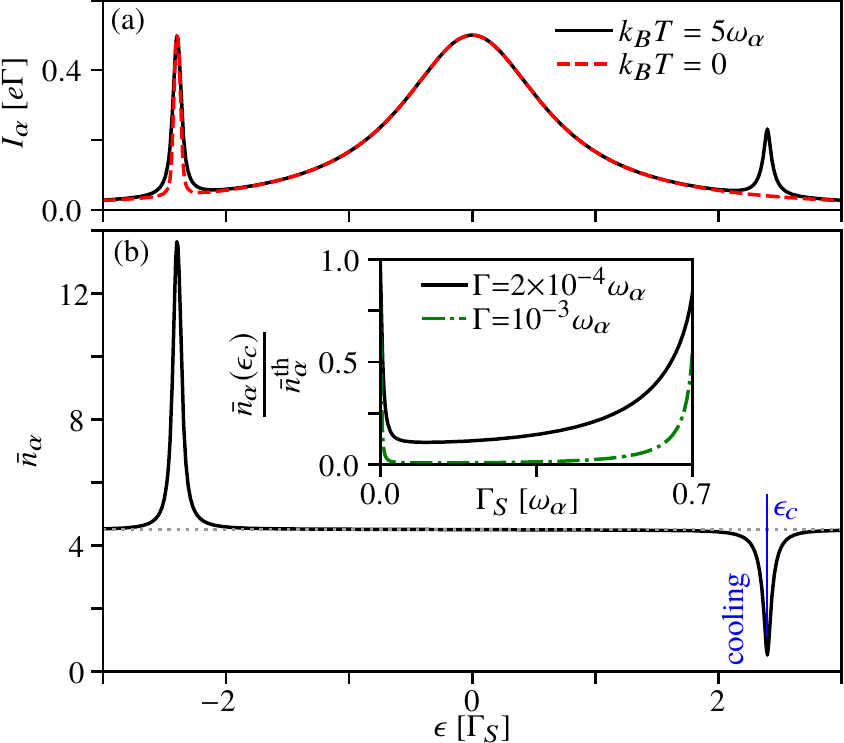}
	\caption{\label{fig:2}
		(a) Current $I_\alpha$ for two identical
		oscillators as a function of the onsite energies $\epsilon$, at zero (dashed line) and finite (solid line) temperature. (b) 
		Average
		photon occupation ${\bar n}_\alpha$ in the $\alpha$-resonator for $k_B T = 5\omega_{\alpha}$. The horizontal dotted line corresponds to the thermal occupation.
		Inset:
		Photon occupation at $\epsilon = \epsilon_c$ as a function of 
		$\Gamma_S$, for
		two different values of $\Gamma$. The curves are rescaled to the 
		thermal
		occupation value.
		Other parameters are $\Gamma=2\times10^{-4}\omega_\alpha,\ \lambda_\alpha=0.02\omega_\alpha,\ Q_\alpha = 10^5,\ \Gamma_S = 0.2\omega_\alpha$.  }
\end{figure}

For $\bar\delta = \omega_L= \omega_R$, one can achieve simultaneous cooling as well as heating of both resonators,
which is already described by the first order terms in $\lambda_\alpha$ of Eq.~\eqref{eq:polaron_transformed_hamiltonian}.  
Here, we consider two identical resonators and tune the dot levels $\epsilon$ 
around the resonance condition $\bar{\delta} \approx \omega_\alpha$, i.e., $\bar{\epsilon}= \pm \sqrt{\smash[b]{\omega_\alpha^2 - 
2\Gamma_S^2}}/2 $.
The effective first-order interaction Hamiltonian reads after a rotating-wave approximation
\begin{equation}
    \label{eq:local_int_hamiltonian}
    H_\mathrm{loc} =
   \sum_\alpha \frac{1}{2} \lambda_\alpha \sin\bar{\theta}\  ( b_\alpha \tau_+ 
   + 
   b_\alpha^\dagger \tau_-),
\end{equation}
as we show in Appendix~\ref{app:app}. The operators $\tau_+
= |+\rangle\langle-|$ and $\tau_- = |- \rangle\langle +|$ describe the hopping between the two-level system formed by the states $|+\rangle$ and $|-\rangle$,
coupled to the modes through a transverse
Jaynes-Cummings-like interaction. The effective coupling is proportional to 
$\sin \bar{ \theta} = \sqrt{2}\Gamma_S/\bar{\delta}$, and, thus, a direct consequence of the nonlocal Andreev reflection.
The effective interaction in Eq.~\eqref{eq:local_int_hamiltonian} coherently mixes the three states $|+,n_L,n_R\rangle, |-, n_L+1, n_R\rangle$, and $|-, n_L, n_R+1\rangle$ which are degenerate for $H_\mathrm{loc}= 0$.
When $|\epsilon| \gtrsim \Gamma_S$, the
hybridization between the charge states is weak.
The sign of $\epsilon$ changes the bare dots' level structure: For 
$\epsilon<0$, $|+\rangle \approx |0\rangle$ and $|-\rangle \approx |S\rangle$, 
whereas for $\epsilon>0$, $|+\rangle \approx |S\rangle$ and $|-\rangle \approx 
|0\rangle$. In the latter case, the chain of transitions 
$|+\rangle \rightarrow |\alpha\sigma\rangle \rightarrow |-\rangle$ is faster 
than the opposite process, see Fig.~\ref{fig:1}(c).
For $\epsilon<0$, energy is pumped into the modes. Conversely, for $\epsilon>0$, we can achieve simultaneous cooling of the resonators.
In Fig.~\ref{fig:2}, we show the stationary electron 
current $I_\alpha$ [calculated using the full Hamiltonian~\eqref{eq:hamiltonian}], together with the average photon number 
$\bar{n}_\alpha = \ev{ b^\dagger_\alpha b_\alpha}$ of the corresponding 
resonator, as a function of $\epsilon$. The broad central resonance of width 
$\Gamma_S$ corresponds to the elastic current contribution mediated by the cross-Andreev reflection. The
additional inelastic peak at negative $\epsilon$ is related to the emission of 
photons in 
both resonators at $\bar{\delta} \approx \omega_\alpha$. At finite 
temperature, a second sideband peak emerges at positive 
$\epsilon$, where the resonators are simultaneously cooled down. The cavities are efficiently cooled into their ground state for a wide range of values of $\Gamma_S$, as can be appreciated in the inset of Fig.~\ref{fig:2}(b). The optimal cooling 
region is due to the interplay between the effective interaction with the
resonator---which vanishes for small $\Gamma_S$---and the hybridization of the
empty and singlet state, which increases as $\epsilon$ approaches the Fermi
level of the superconductor and reduces the asymmetry of the transitions $|\pm\rangle \leftrightarrow |\alpha\sigma\rangle$.
%
%
%
\section{Nonlocal photon transfer}
\label{sec:nonlocal}
\begin{figure}[t]
	\includegraphics[]{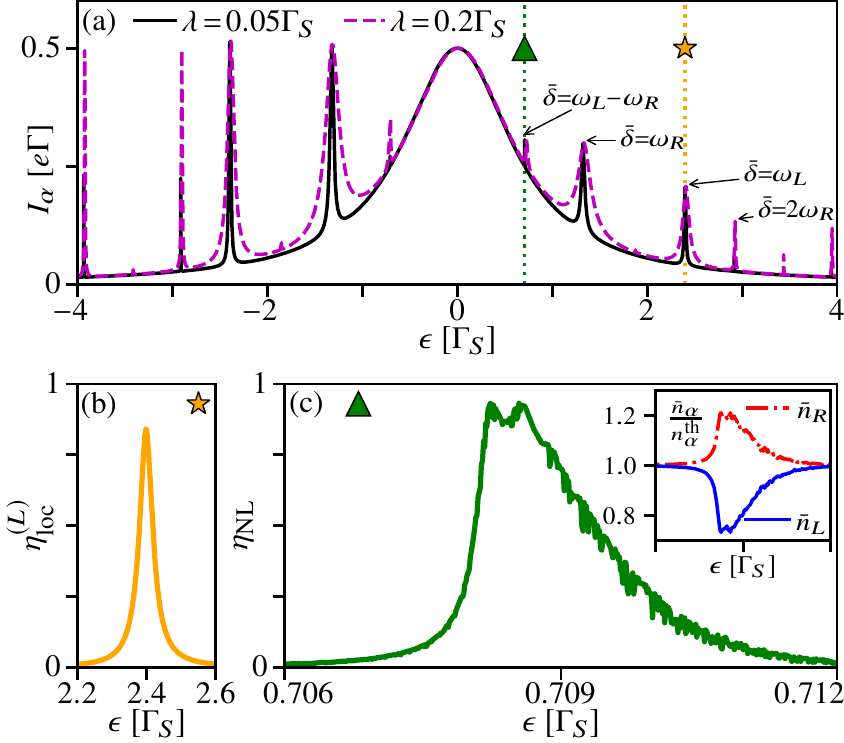}
	\caption{\label{fig:3} (a) Current $I_\alpha$ through lead $\alpha$ as a function of
	the onsite energies $\epsilon$, for two different values of $\lambda\equiv
	\lambda_L = \lambda_R$. The arrows indicate resonances according to Eq.~\eqref{eq:resonances}.
	(b) Local cooling efficiency for the left mode, around $\bar{\delta} \approx
	\omega_L$.
	(c) Photon transfer efficiency around $\bar{\delta} \approx \omega_L -
	\omega_R$.
	Inset: Average cavity photon number, normalized to the thermal occupation.
	Parameters are $\Gamma = 10^{-4}\Gamma_S,\ \omega_L = 5\Gamma_S,\ \omega_R =
	3\Gamma_S,\ Q_L = Q_R = 10^5,\ T = 5\Gamma_S$.}
\end{figure}
By keeping terms up to second order in $\lambda_\alpha$ in 
Eq.~\eqref{eq:polaron_transformed_hamiltonian}, we can describe the resonances 
around 
$\bar{\delta} = \omega_L - \omega_R$ and $\bar{\delta} = 
\omega_L + 
\omega_R$. Assuming without loss of
generality $\omega_L > \omega_R$, a rotating-wave approximation yields the 
effective interaction terms $H^{(-)}_\mathrm{NL} =
\lambda_\mathrm{NL} (b^\dagger_L b_R \tau_- + \mathrm{H.c.})$ for $\bar{\delta 
}
\approx \omega_L- \omega_R$, and
$H^{(+)}_\mathrm{NL} = \lambda_\mathrm{NL} (b_L b_R \tau_+ +
\mathrm{H.c.})$ for $\bar{\delta} \approx \omega_L+ \omega_R$, see Appendix~\ref{app:app}.
These 
terms show that the two resonators become indirectly coupled through the 
charge states, with the strength
\begin{equation}
	\label{eq:nonlocal_coupling}
    \lambda_\mathrm{NL} = \frac{\Gamma_S \lambda_L \lambda_R}{\sqrt{2} 
    \omega_L \omega_R} \cos \bar{\theta}.
\end{equation}
We remark that this interaction is, as well, purely nonlocal.
$H^{(+)}_\mathrm{NL}$ describes the hybridization of the states in the subspace $|+,
n_L-1, n_R -1\rangle$ with $|-, n_L, n_R\rangle$, through which photons at different frequencies are
simultaneously absorbed (emitted) from (into) both cavities. Conversely, the term  $H^{(-)}_\mathrm{NL}$ describes processes by which the superconductor
mediates a coherent transfer of photons between the resonators, by coupling
the subspaces $|+, n_L-1, n_R+1 \rangle$ and $|-, n_L, n_R  \rangle$, see 
Fig.~\ref{fig:1}(d).  Notice
that this effect vanishes if the two resonators are of the same frequency, as it would require 
$\bar{\delta} = 0$
and, thus, $\Gamma_S = 0$.
In Fig.~\ref{fig:3}(a), we report the electronic current, again calculated with the full interaction,
assuming two different resonator frequencies. In addition
to the sideband peaks close to $\bar{\delta}=\omega_L$ and 
$\bar{\delta}=\omega_R$, we can identify
higher-order multiphoton resonances (e.g., $\bar{\delta}=2\omega_R$, where the 
cooling
cycle involves the absorption of two photons from the same cavity) which can be 
described in a similar
way with a rotating-wave approximation, see Appendix~\ref{app:app}. Moreover,
we observe the second-order peaks described by $H_\mathrm{NL}^{(\pm)}$ which are
responsible for processes involving both resonators. The inset of
Fig.~\ref{fig:3}(c) reports the average occupation of the resonators in the
vicinity of the resonance $\bar{\delta} = \omega_L - \omega_R$, where the right
mode is heated and the left one is cooled. The shape of these resonances 
differs from the first-order peaks (which are well approximated by  
Lorentzians): We show in Appendix~\ref{app:app} how the second-order Hamiltonian contains indeed an additional 
term proportional to $\sin (\bar{\theta}) (2n_\alpha + 1) \tau_z$,
which 
causes both 
a small frequency shift for each resonator (yielding a double-peak structure) and a small renormalization of the 
splitting $\bar{\delta}$ between the Andreev bound states. Nevertheless, this 
corrections do not alter the main physics captured 
by $H_{\mathrm{NL}}^{(-)}$.

\section{Heat transfer and efficiency}
\label{sec:efficiency}
To quantify the performance of both cooling and nonlocal photon transfer, we 
calculate the stationary heat 
current~\cite{Erdman2017,Erdman2018,Benenti2017}
\begin{align}
	\label{eq:energy_current}
	\dot{E}^\mathrm{ph}_\alpha &= \sum_{i, j, s} E_{ij} w_{\mathrm{ph}, j
	\leftarrow i}^{\alpha, s} P^{\mathrm{st}}_i
\end{align}
flowing from the bosonic
reservoir $\alpha$ to the corresponding resonator.
It is negative (positive) when the resonator is
cooled (heated), and vanishes for an oscillator in
thermal equilibrium. As a figure of merit for local cooling, we can 
estimate the number of bosonic quanta subtracted from the resonator on average 
per 
unit time, and compare it to the rate at which Cooper pairs are injected into 
the system. The latter rate is given by $|I_S|/2e$ with $I_S = -(I_L + 
I_R)$ 
being the Andreev current through the superconductor found from current conservation. Consequently, the local 
cooling 
efficiency around $\bar{\delta} = \omega_\alpha$ can be defined as 
$\eta^{(\alpha)}_{\mathrm{loc}} = 
\frac{2e|\dot{E}_\alpha|}{|I_S| \omega_\alpha}$. Similarly, around 
$\bar{\delta} = \omega_L- \omega_R$, we define the heat transfer efficiency
\begin{equation}
\label{eq:nonlocal_efficiency}
\eta_{\mathrm{NL}} = \frac{2e|\dot{E}_L - \dot{E}_R|}{|I_S| (\omega_L- 
\omega_R)}.
\end{equation}
Figures~\ref{fig:3}(b) and (c) show $\eta^{(L)}_\mathrm{loc}$ and 
$\eta_\mathrm{NL}$, respectively, as a function of $\epsilon$ close to the corresponding 
resonances. In both cases, we obtain high efficiencies close to 90\%: 
Approximately one photon is absorbed from each cavity (local cooling) or 
transferred from the left to the right cavity (nonlocal transfer) per Cooper 
pair. 
The efficiency is essentially limited by two factors: (i) an elastic
contribution to the current [the broad resonance of linewidth $\propto\!\Gamma_S$ in
Figs.~\ref{fig:2}(a) and \ref{fig:3}(a)] where electrons flow without exchanging 
energy with the cavities; (ii) a finite 
fraction of the injected electrons acting against the dominant 
process (cooling or photon transfer), as illustrated by the dashed blue arrows 
in Fig.~\ref{fig:1}(d). Both processes augment with increasing $\Gamma_S$ and are a byproduct of 
the finite hybridization between the empty and the singlet state which, however, is crucial 
for achieving a nonzero efficiency.

\section{Conclusions}
\label{sec:conclusions}
We have analyzed a CPS in a double-quantum-dot setup, with local
charge couplings to two resonators. We have demonstrated that Cooper-pair splitting can generate a nonlocal transfer of photons and heat from one
oscillator to the other, resulting in a stationary energy flow. Such energy flows can also be channeled to cool or heat locally a single cavity. Hence, our system constitutes a versatile tool to fully inspect heat exchange mechanisms in hybrid systems, and is a testbed for quantum thermodynamics investigations involving both electronic and bosonic degrees of freedom.
Due to the single-photon nature of the coherent interactions, this can also be extended to achieve
few-phonon control and manipulation,~\cite{Okamoto2013,Zhu2017} e.g., by 
implementing time-dependent protocols for the dots' gate voltages to tune 
dynamically the strength of the nonlocal features. Further practical 
applications include high-efficiency nanoscale heat pumps and cooling devices 
for nanoresonators. 

A discussion on the experimental feasibility of our setup is in order. 
For 
single quantum dots coupled to microwave resonators, $\lambda_\alpha/(2\pi)$ 
can 
reach 100 MHz, with resonators of quality factors 
$Q\!\!\sim\!\!10^4$ and frequencies 
$\omega_\alpha/(2\pi)\!\!\sim\!\!7$ GHz.~\cite{Bruhat2016,Stockklauser2017} For mechanical 
resonators, 
coupling 
strengths of $\lambda_\alpha/(2\pi)\!\!\sim\!\! 100$ kHz for frequencies of 
order 
$\omega_\alpha/(2\pi)\!\!\sim\!\!1$ MHz and larger quality factors up to $10^5 
\!\!-\!\!10^6$ have been reported.~\cite{Okazaki2016} In a double-quantum-dot 
Cooper-pair splitter setup, the cross-Andreev reflection rate is approximately $\Gamma_S\!\!\lesssim\!\!\sqrt{\Gamma_{SL} \Gamma_{SR}}$, when the 
distance between the dots is much shorter than the coherence length in the 
superconducting contact.\cite{Hussein2019} Here, $\Gamma_{S\alpha}$ is the 
local Andreev 
reflection rate which can reach several tens of $\mu$eV, becoming comparable 
to the typical microwave resonator frequencies (thus allowing $\Gamma_S \lesssim 
\omega_\alpha$) while being order of 
magnitudes lower than the superconducting gap $\Delta$.\cite{Gramich2015} Therefore, 
the regime of parameters we considered lies within the range of 
state-of-the-art technological capabilities. 
Moreover,  experiments involving Cooper-pair 
splitters~\cite{Hofstetter2009, Herrmann2010, Hofstetter2011, Das2012, 
Schindele2012,
Schindele2014, Fulop2014, Fulop2015, Tan2015, Borzenets2016, Baba2018} or
mesoscopic cQED devices with microwave 
cavities~\cite{Frey2012,Delbecq2013,Deng2015,Stockklauser2017,Viennot2015,Mi2017,
	Li2018,Cubaynes2019} and mechanical 
	resonators~\cite{Benyamini2014,Okazaki2016,Deng2016,Urgell2019} are of 
	appealing and growing interest, and therefore promising candidates for the 
	implementation of the system described here.

\begin{acknowledgments}
This research was supported by the German Excellence Initiative through the
Zukunftskolleg and by the Deutsche Forschungsgemeinschaft through the SFB 767.
R.H. acknowledges financial support from the Carl-Zeiss-Stiftung.
\end{acknowledgments}
\begin{appendices}
\appendix
\section{Polaron-transformed Hamiltonian and effective nonlocal	interaction} 
	\label{app:app}
	
	We report here the derivation of the effective interactions
	that explain the local cooling or heating, and the nonlocal photon transfer
	mechanisms. The starting point is the polaron-transformed Hamiltonian given
	in Eq.~\eqref{eq:polaron_transformed_hamiltonian} of the main text.
	For small coupling strengths $\lambda_\alpha \ll \omega_{\alpha}$, we expand the operators $X$
	and $X^\dagger$ up to second order in $\lambda_\alpha$. The dots-cavities 
	interaction term is
	\begin{equation}
		\label{eq:interaction_second_order}
		H_{\mathrm{int}} = -\frac{\Gamma_S}{\sqrt{2}} \left[ i \sigma_y \Pi
		 + \sigma_x \left(1 +
		\frac{\Pi^2}{2} \right) \right] + \mathcal{O}(\Pi^3),
	\end{equation}
	with $i\Pi = \sum_{\alpha} i\Pi_\alpha$ the generalized total momentum, 
	$\sigma_x = |0 \rangle \langle S| +
	\text{H.c.}$  and $\sigma_y = -i|0 \rangle \langle S| + \text{H.c.}$ The
	$\sigma_x$\nobreakdash-term describes tunneling between the empty and
	the singlet state due to the superconductor, and is already present in
	Hamiltonian~\eqref{eq:hamiltonian} of the main text. Diagonalizing the bare electronic part leads to the hybridized charge
	states
	\begin{align}
		\label{eq:hybridized_states}
		|+\rangle &= \cos \left(\frac{\bar{\theta}}{2} \right)|0 \rangle +   
		\sin \left(\frac{\bar{\theta}}{2} \right)|S \rangle, \\
		|-\rangle &= -\sin \left(\frac{\bar{\theta}}{2} \right)|0 \rangle +   
		\cos \left(\frac{\bar{\theta}}{2} \right)|S \rangle,	
	\end{align}
	 with the mixing angle
	$\bar{\theta}$ and the energy splitting $\bar{\delta}$ defined in the main
	text. By introducing the Pauli matrices $\tau_x = \tau_+ + \tau_-$, $\tau_y = -i (\tau_+ - \tau_-)$, $\tau_z = [\tau_+, \tau_-]$ with $\tau_+= |+\rangle\langle-|$ and $\tau_- = |- \rangle\langle +|$, we can express the Hamiltonian~\eqref{eq:polaron_transformed_hamiltonian} of main text to second order by
\begin{equation}
	\label{eq:hamiltonian_second_order}
	\begin{split}
	\bar{H} &= \sum_{\alpha\sigma} \bar{\epsilon}_\alpha N_{\alpha \sigma} +
	\frac{\bar{\delta}}{2} \tau_z + \sum_\alpha \omega_\alpha b^\dagger_\alpha
	b_\alpha \\
	& - \frac{\Gamma_S}{2\sqrt{2}} \left[  2 i\tau_y \Pi + (\sin 
	\bar{\theta}\ 
	\tau_z + \cos
	\bar{\theta}\  \tau_x) \Pi^2 \right] + \mathcal{O}(\Pi^3).
	\end{split}
\end{equation}
	We now move to the interaction picture with respect to the noninteracting Hamiltonian $H_0
	= \sum_{\alpha\sigma} \bar{\epsilon}_\alpha N_{\alpha \sigma} + \frac{\bar{\delta}}{2} \tau_z 
	+
	\sum_\alpha \omega_\alpha b^\dagger_\alpha b_\alpha$. By recalling the definition of $\Pi$, we obtain in the interaction picture the Hamiltonian
	\begin{widetext}
		\begin{equation}
		\label{eq:hamiltonian_second_order_expanded}
		\begin{split}
			H_\text{int}(t) &= -\sum_{\alpha} \frac{\lambda_\alpha
			\Gamma_S}{\omega_\alpha \sqrt{2}} \left(e^{i\omega_\alpha t} 
			b_\alpha^\dagger - e^{-i\omega_\alpha t}
			b_\alpha  \right)
			\left(e^{i\bar{\delta} t}\tau_+ - e^{-i \bar{\delta} t} \tau_- \right)  \\
			&- \frac{\Gamma_S \lambda_L \lambda_R}{ \sqrt{2}
				\omega_L \omega_R} \left[e^{i \Omega t} b_L^\dagger
				b_R^\dagger + e^{-i \Omega t} b_L
				b_R - e^{i (\Delta \omega)  t} b_L^\dagger
				b_R - e^{-i (\Delta \omega)  t} b_L
				b_R^\dagger   \right]
			\left[ \sin( \bar{\theta}) \tau_z + \cos (\bar{\theta} )(e^{i\bar{\delta} t} \tau_+ +
			e^{-i \bar{\delta} t} \tau_-)\right] \\
			&-  \sum_{\alpha} \frac{\Gamma_S \lambda_\alpha^2}{2 \sqrt{2}
			\omega_\alpha^2} \left[e^{2i\omega_\alpha t}(b_\alpha^\dagger)^2 +
			e^{-2i\omega_\alpha t} b_\alpha^2 - 2 b^\dagger_\alpha b_\alpha- 1  \right] \left[
			\sin
			(\bar{\theta}) \tau_z + \cos
		(	\bar{\theta}) (e^{i\bar{\delta} t} \tau_+ +
		e^{-i \bar{\delta} t} \tau_-)\right] + \mathcal{O}(\lambda_\alpha^3/\omega_{\alpha}^3).
		\end{split}
		\end{equation}
	\end{widetext}
	Here, we have introduced $\Omega = \omega_L+ \omega_R$ and $\Delta\omega = \omega_L 
	-
	\omega_R$. Hamiltonian~\eqref{eq:hamiltonian_second_order_expanded} 
	contains all the terms that lead to cooling, heating, and nonlocal photon 
	transfer. To isolate these features, we will focus on the
	relevant resonances $\bar{\delta} \approx \omega_\alpha$, $\bar{\delta} \approx
	\Omega$, and $\bar{\delta} \approx \Delta\omega$. 
	First, let us consider two identical resonators of frequency $\omega_\alpha = \omega$ and tune 
	$\epsilon$ such that 
	$\bar{\delta} = \omega$. Notice that this can be fulfilled by two values of $\epsilon$, of opposite sign. In the following, we restrict 
	Eq.~\eqref{eq:hamiltonian_second_order_expanded} to first order in 
	$\lambda_\alpha$, and then discard the fast-oscillating terms by 
	performing a standard rotating-wave approximation (RWA). Thus, we obtain the time-independent interaction Hamiltonian given by Eq.~\eqref{eq:local_int_hamiltonian} in the main text,
	\begin{equation}
		\label{eq:hamiltonian_first_order}
		H^{\bar{\delta}=\omega}_\mathrm{RWA} = \sum_{\alpha} \frac{1}{2} 
		\lambda_\alpha 
		\sin(\bar{\theta})\  ( b_\alpha \tau_+ 
		   + 
		   b_\alpha^\dagger \tau_-).
	\end{equation}
	We have used here the resonance condition $\omega = \bar{\delta}$ and 
	the relation $\sin\bar{\theta} = \sqrt{2}\Gamma_S/\bar{\delta}$. 
	
	Let us now consider the nonlocal resonance, $\bar{\delta}=\Delta\omega$.
    A peculiarity is here, that we have to go to second order in $\lambda_\alpha$,
    since the first-order terms become in the RWA fast rotating and, thus, average to zero. The 
	corresponding effective Hamiltonian reads
	\begin{equation}
		\label{eq:hamiltonian_second_order_diff}
		H^{\bar{\delta} = \Delta\omega}_{\mathrm{RWA}} =\sum_\alpha 
		\frac{\Gamma_S\lambda_\alpha^2}{2\sqrt{2}\omega_\alpha^2}(2 n_\alpha + 
		1)\sin\bar{\theta}
		 \tau_z + 
		 \lambda_{\mathrm{NL}} (b_L^\dagger b_R 
		\tau_- + 
		\mathrm{H.c.}),
	\end{equation}
	with $n_\alpha = b_\alpha^\dagger b_\alpha$ the photon number operator, 
	and $\lambda_{\mathrm{NL}}$ stated in Eq.~\eqref{eq:nonlocal_coupling} of 
	the main text. The second term corresponds to the interaction $H_\mathrm{NL}^{(-)}$ (main text), and is responsible for the coherent transfer of photons between the cavities, leading to a stationary energy flow. The first term in Eq.~\eqref{eq:hamiltonian_second_order_diff} proportional to $n_\alpha\tau_z$ can be seen as a dispersive shift of the cavity frequencies, which depends on the Andreev bound state. As the quantities reported in Fig.~\ref{fig:3} of the main text are averages calculated from the density matrix, this translates into a fine double-peak structure of the nonlocal resonance, see Fig.~\ref{fig:3}(c) of the main text. Further, the additional term proportional to $\tau_z$ 
	renormalizes the level splitting $\bar{\delta}$ and, therewith, the resonance condition, $\bar{\delta} = \Delta \omega$. 

	Considering the condition $\bar{ \delta} = \Omega$, we obtain the effective RWA Hamiltonian
	\begin{equation}
		\label{eq:hamiltonian_second_order_sum}
		H^{\bar{\delta} = \Omega}_{\mathrm{RWA}} =\sum_\alpha 
		\frac{\Gamma_S\lambda_\alpha^2}{2\sqrt{2}\omega_\alpha^2}(2 n_\alpha + 
		1)\sin\bar{\theta}
		 \tau_z + 
		 \lambda_{\mathrm{NL}} (b_L^\dagger b_R^\dagger 
		\tau_- + 
		\mathrm{H.c.}).
	\end{equation}
	Here, the relevant interaction ($H^{(+)}_{\mathrm{NL}}$ of the main text) describes absorption (and emission) from \emph{both} cavities simultaneously while flipping the Andreev state. So, this second-order effect may entail simultaneous cooling, $\epsilon>0$, and heating, $\epsilon<0$, of both cavities.
	
	From the last line of Eq.~\eqref{eq:hamiltonian_second_order_expanded}, 
    one can infer an effective RWA Hamiltonian governing the resonance condition $\bar{ \delta} \approx 2 \omega_{\alpha}$. It is similar to Eq.~\eqref{eq:hamiltonian_first_order}, but involves absorption and emission of two photons from the same cavity. Indeed, this two-photon resonance is also observable in Fig.~\ref{fig:3}(a) of the main text and yields cavity cooling for $\epsilon>0$ and heating for $\epsilon<0$, respectively. 
	
	By including terms up to $n$-th order in $\Pi$ in Eq.~\eqref{eq:hamiltonian_second_order}, one obtains terms  $(b_\alpha)^n$ and $(b^\dagger_\alpha)^n$, which, after moving to the interaction picture and performing a suitable RWA, will yield $n$-photon local absorption/emission processes. The expansion contains also terms of the form $(b_\alpha^\dagger)^{p} (b_{\bar{\alpha}})^{q}$ and $(b_\alpha^\dagger)^{p} (b^\dagger_{\bar{\alpha}})^{q}$ together with their Hermitian conjugates, with $p +q = n$ ($\bar{\alpha} = R$ if $\alpha = L$ and vice versa). The former terms describe the coherent transfer of $|p- q|$ photons between the cavities, while the latter describes coherent emission and re-absorption of $p$ and $q$ photons from the $\alpha$ and $\bar{\alpha}$ cavity, respectively.  The general (approximate) resonance condition thus reads \mbox{$\bar{ \delta} \approx |p \omega_L \pm q \omega_R|$}, stated in Eq.~\eqref{eq:resonances} in main text. If either $p$ or $q$ is zero, the resonance corresponds to local cooling/heating of the cavities.

\end{appendices}

%



\end{document}